\documentclass[]{aipproc}

\layoutstyle{8x11double}

\begin{document}

\title[Radio/Sub-mm/X-ray Studies of GRB Hosts]{Radio, Sub-mm, and
X-ray Studies of Gamma-Ray Burst Host Galaxies}

\author{E. Berger}{
address={Palomar Observatory, California Institute of Technology 105-24, Pasadena, CA 91125},
email={ejb@astro.caltech.edu},
}

\begin{abstract}
The study of gamma-ray burst (GRB) host galaxies in the radio, sub-mm,
and X-ray wavelength regimes began only recently, in contrast to optical
studies.  This is mainly due to the long timescale on which the radio
afterglow emission decays, and to the intrinsic faintness of radio
emission from star-forming galaxies at $z\sim 1$, as well as source
confusion in sub-mm observations; X-ray observations of GRB hosts have
simply not been attempted yet.  Despite these difficulties, we have
recently made the first detections of radio and sub-mm emission from
the host galaxies of GRB\,980703 and GRB\,010222, respectively, using
the VLA and the SCUBA instrument on JCMT.  In both cases we find that
the inferred star formation rates ($\sim 500$ M$_\odot$) and
bolometric luminosities (${\rm few}\times 10^{12}$ L$_\odot$) indicate
that these galaxies are possibly analogous to the local population of
Ultra-Luminous Infrared Galaxies (ULIRGs) undergoing a starburst.
However, there is a modest probability that the observed emission is
due to AGN activity rather than star formation, thus requiring
observations with Chandra or XMM.  The sample of GRB hosts
offers a number of unique advantages to the broader question of the
evolution of galaxies and star formation from high redshift to the
present time since: (i) GRBs trace massive stars, (ii) are detectable
to high redshifts, and (iii) have immense dust penetrating power.
Therefore, radio/sub-mm/X-ray observations of GRB hosts can
potentially provide crucial information both on the nature of the GRB
host galaxies, and on the history of star formation.  
\end{abstract}

\maketitle

\begin{section}{Background: The Star Formation History of the Universe}

The formation and evolution of galaxies from high redshift to the current
time is a major focus of modern cosmology. This involves mapping the
conversion of gas into stars, and the associated buildup
of heavy elements. The former is succinctly parameterized by using the
light from massive stars as a surrogate for the star-formation rate (SFR;
see \cite{mfd+96}) and the latter has focused on
studies of the intergalactic medium (e.g. \cite{rauch98})
and damped-Ly$\alpha$ systems (e.g. \cite{sw00}).  GRBs can
contribute to both these important
areas through multi-wavelength studies of their host galaxies, and
through absorption spectroscopy of their optical afterglows.  Here I
address the first issue.

Adelberger \&\ Steidel (2000)\citep{as00} provide a balanced review
of the various diagnostics to measure the evolution of SFR: optical
(rest-frame UV), far-infrared (FIR), sub-millimeter, and decimeter radio
measurements.  Each of these techniques has its distinct strengths
{\it as well as} weaknesses.  For example, radio measurements offer superb
astrometry but the current VLA sensitivity is only able to identify the
tip of the star-formation iceberg \cite{rkf+98}.  Optical surveys
offer the highest 
sensitivity but are vulnerable to dust extinction and may well miss
galaxies forming stars at the most prodigious rate.  The FIR and sub-mm
approach has maximal sensitivity to dusty galaxies, but it lacks
astrometric precision thereby creating a non-trivial bottleneck of
requiring detection at other wavelengths; in particular detection by
the VLA. 

The principal question is the following: Given that each of these
techniques provides a restricted view of the cosmic SFR history, and that
the optical/UV technique has provided by far the largest sample, can we
conclude that optical/UV studies have more or less accounted for the bulk
of the cosmic star formation?

Adelberger \&\ Steidel (2000) seem to think so; a number
of other authors, especially those using long wavelength techniques
(e.g. \cite{bcr00,prb+00})
are of the opinion that this issue is not settled.  In the near term, we
have reached a stalemate since sub-mm surveys
have reached  the confusion limit.  SIRTF, through its IRAC survey,
and ALMA, with its unprecedented $\mu$Jy sensitivity
will certainly contribute to this critical sub-field of modern
astronomy.  Nonetheless, since the bulk of the stellar energy is
effectively radiated in the FIR band, all these techniques require
significant extrapolation to measure the true power radiated by
galaxies. In the distant future, one can envisage FIR interferometers
as providing the most decisive picture of the cosmic
evolution of stellar energy density. 

\end{section}

\begin{section}{GRB Host Galaxies: Strengths}

The host galaxies of GRBs offer a unique perspective
into the SFR history of the Universe for the following reasons:
\begin{enumerate}

\item The existing data show excellent circumstantial evidence
linking GRBs to massive stars (e.g. \cite{bkd01}). More to the point,
every well-studied GRB so far has been identified with a
host galaxy (Fig.~\ref{fig:opt}).

\item GRBs are so bright that they are detectable to redshifts $>20$
(should they exist; \cite{lr00}).  Thanks to the
broad-band afterglow spectrum (X-ray through decimeter radio)
not only can the host be accurately localized, but the 
redshift can also be obtained.  Usually $z$ is obtained
via optical spectroscopy, though with GRB\,990705
\cite{afv+00} we now have a redshift from X-ray spectroscopy
of the afterglow.  

\item The immense dust-penetrating power of GRBs (only limited by
Compton thick column densities) results in a sample of galaxies that
is independent of the global dust properties.

\item Again thanks to the afterglow, a GRB host galaxy 
need only be detected via imaging (with $z$ via absorption
spectroscopy) and this has resulted in the faintest luminosity
distribution of star-forming galaxies with some hosts 12
magnitudes below $L_*$ (Fig.~\ref{fig:opt}).  In contrast, current
state of the art optical/UV/NIR surveys reach $\sim L_*$.

\end{enumerate}
\begin{figure}
\resizebox{0.48\textwidth}{!}{\includegraphics{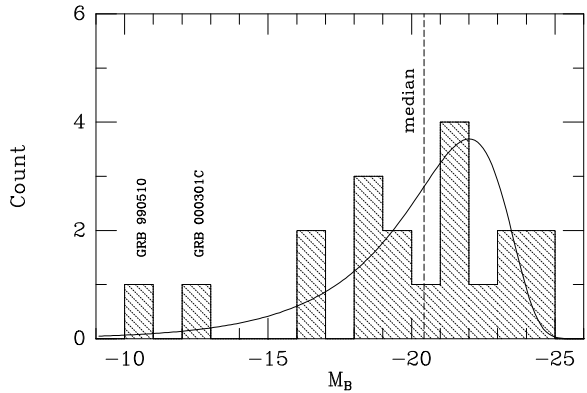}}
\caption{Histogram of estimated absolute B-band magnitudes of GRB
host galaxies with known redshifts.  These rest-frame magnitudes were
computed from the observed R-band magnitudes by approximating the
galaxy spectra as $f_\nu\propto \nu^{-1}$.  The sample median is
$M_{\rm B}=-20.4$ mag.  The solid curve is a heuristic model
representing a luminosity-weighted Schechter function with $M_*=-23$
mag and $\alpha=-1.6$.}
\label{fig:opt}
\end{figure}

\end{section}

\begin{section}{GRB Host Galaxies: Possible Drawbacks}

It is clear that the sample of GRB hosts offers powerful
diagnostics in our quest to decipher the SFR history of the
Universe.  However, it has two limitations.  First, we have
to assume that the GRB rate is linearly proportional to SFR.  The
circumstantial evidence for the association of GRBs with massive
stars, and hence SFR is good \cite{pgg+00,bkf01,bkd01,fbm+01,rei01}.
Second, the GRB sample is quite small, especially when
compared to the optical/UV sample.

However, the first problem is not as severe as one may think at first
glance.  All techniques used so far -- optical/UV, ISOCAM, sub-mm and
decimeter -- require large extrapolations (and implicitly, constancy of
spectral energy densities) to obtain the bolometric power.
Converting this uncertain bolometric estimate to SFR requires detailed
assumptions of the IMF of stars and the distribution of ISM in these
distant galaxies.  
The severity of the second problem diminishes when one realizes that
the number of {\it securely} identified
sub-mm galaxies is, as of August 2001, only four! (M. Longair, talk at
ESO Lighthouse conference).

\end{section}

\begin{section}{The Origin of Radio and Sub-mm Emission from
Galaxies}

Having argued that GRBs offer a unique perspective into the
cosmic star formation, I now provide a short overview of the
underlying sources and emission mechanisms of radio and sub-mm
emission from galaxies. 
The radio luminosity from star-forming galaxies is a combination of
synchrotron and thermal emission components, both directly related to
the formation rate of massive stars via simple relationships
\cite{c92}.  This is simply due to the fact that radio
synchrotron emission comes from electrons accelerated in supernova
shocks, the end products of massive stars, and thermal emission comes
from HII regions and is dominated by the most luminous (i.e. massive)
stars.  In addition, since the lifetime of massive stars is $\sim
10^7$ years, and the lifetime of the synchrotron emitting electrons is 
$\sim 10^8$ years, radio emission traces the instantaneous SFR
\cite{c92}.  

Similarly, sub-mm (and FIR) emission traces star
formation since it arises from star-light reprocessed by dust.  In
this case too the massive stellar population dominates the power
output in the host, and therefore the amount of reprocessed radiation.
Since the emission in the radio and sub-mm regimes is a tracer of the
massive stellar population, it is not surprising that there is a
simple relation between the radio and sub-mm luminosities of star
forming galaxies.  It turns out that this relation is sensitively
dependent on redshift \cite{cy99} \cite{cy00} \cite{dce00}.  

One complication to the preceding discussion is the possibility of 
emission from an obscured AGN, which will contribute to both the radio
and sub-mm luminosities of the host.  Observations with X-ray
satellites can provide an estimate of the fraction of emission (if
any) that arises from an active nucleus.

\end{section}

\begin{section}{Recent Detections: GRB\,980703 and GRB\,010222}

Recently, we have detected the host galaxy of GRB\,980703 in
the radio \cite{bkf01}, and the host of GRB\,010222 in the
sub-mm \cite{fbm+01}.  Fig.~\ref{fig:980703lc} shows the light-curve of 
the 8.46 GHz emission from GRB\,980703; the flattening at
$t>350$ days can only be explained in terms of host emission.
We detect similar levels of emission at 1.43 and 4.86 GHz.  At a
redshift of $z=0.966$ these flux levels translate to an emitted
luminosity at 1.43 GHz of $L_{\rm em}(1.43)\approx 4.7\times 10^{30}$
erg sec$^{-1}$ Hz$^{-1}$.  

What can we learn about the host galaxy of GRB\,980703 from
the observed radio luminosity?  First, the emitted 1.43 GHz luminosity
immediately translates to a formation rate of stars more massive than
5 M$_\odot$, ${\rm SFR(M>5M_\odot)}\approx 90$ M$_\odot$/yr, and a
total star formation rate (using the Salpeter IMF) of $\approx 500$
M$_\odot$/yr.  Second, based on this luminosity and the radio/sub-mm
relation, we find that this galaxy is an Ultra-Luminous Infra-Red
Galaxy (ULIRG; see \cite{sm96}), with $L_{\rm
FIR}\approx 10^{12}$ L$_\odot$.  A comparison to the properties of
radio-selected galaxies at $z\sim 1$ from a survey of the HDF
\cite{hpw+00}, shows that the host of GRB\,980703 is by no means
an unusual galaxy.  On the other hand, a comparison to the
optically-derived SFR ($\sim 20$ M$_\odot$/yr; \cite{dkb+98}) shows
that most of the star formation in this galaxy is obscured. 

An alternative explanation for the radio emission is that it originates
from an AGN. Surveys of
the Hubble Deep Field (HDF), its flanking fields, and the Small
Selected Area 13 (SSA13) have shown that approximately 20\% of the radio sources
are AGN \cite{rfk+99,r00,bcr00}.

We consider the AGN hypothesis unlikely based on 
optical spectroscopy.  Optical spectra of the source obtained by
Djorgovski et al. (1998) show no evidence for an unobscured AGN:
high-ionization lines such as Mg II $\lambda 2799$, [NeV]$\lambda
3346$, and [NeV]$\lambda 3426$ are absent, and the [OIII]$\lambda
4959$ to H$\beta$ ratio is approximately 0.4, much lower than ${\rm
  [OIII]/H}\beta>1.3$ for AGN \citep{rtt97}.  
In addition, AGN have redder colors for
similar [OII] EW, relative to normal galaxies \cite{k92}.  Using the
spectrum of GRB\,980703 we evaluate the color index,
$(41-50)\equiv 2.5{\rm log}[f_\nu(5000{\rm \AA})/f_\nu(4100{\rm
\AA})]\approx 0\pm 0.1$; an AGN with the
same [OII] EW would have a value $> 0.3$ \citep{k92}.  

However, it is not possible to rule out the existence of an obscured
AGN.  Future observations with XMM will probe this possibility
directly. 

The high resolution afforded by VLA observations has shown
that the GRB-host offset of GRB\,980703 is negligible (see
Fig.~\ref{fig:offset}), indicating that the burst most
probably took place within a nuclear starburst; in this
case the resolution of the VLA allows a better offset determination
than HST observations \cite{bkd01}.  The nuclear starburst origin
lends strong support to the collapsar model of GRBs.

The sub-mm detection of the host of GRB\,010222 paints a
similar picture (Fig.~\ref{fig:010222sp}).  The implied SFR is close
to 1000 $M_\odot$/yr, and the FIR luminosity clearly indicates that
this host is also a ULIRG \cite{fbm+01}.  

\begin{figure}
\resizebox{0.48\textwidth}{!}{\includegraphics{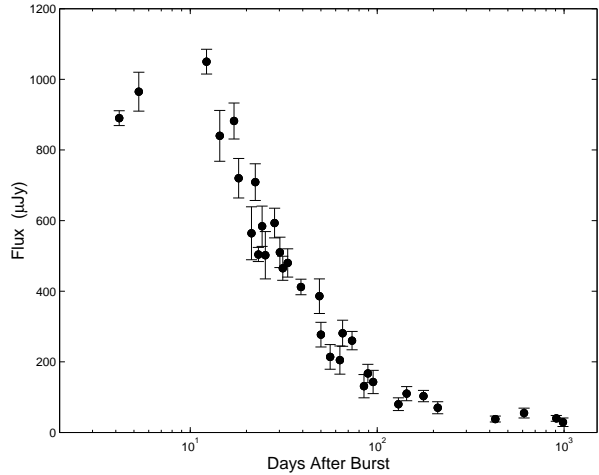}}
\caption{Radio light-curve at 8.46 GHz, showing the customary
initial rise followed by decay of the afterglow of GRB\,980703 (Berger
et al. 2001).  Observations from $\sim 100-300$ days after the burst
already show signs of flattening, due to the flux contribution from
the host, while observations from $t>350$ days directly probe the
emission of the host.  On these timescales, the afterglow contribution
is negligible.} 
\label{fig:980703lc}
\end{figure}

\begin{figure}
\resizebox{0.48\textwidth}{!}{\includegraphics{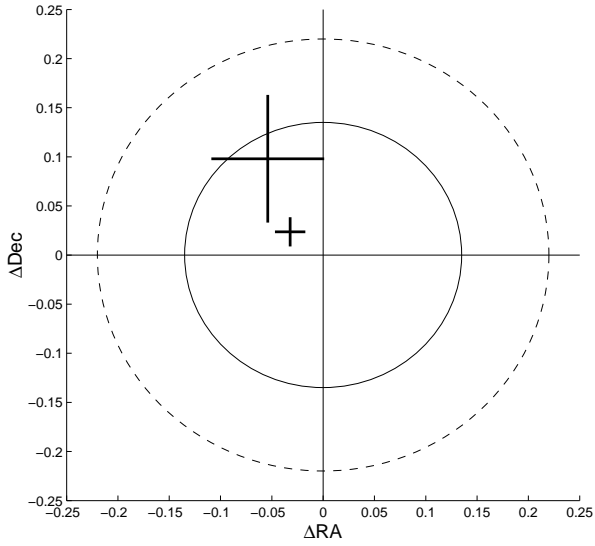}}
\caption{The weighted average GRB-host offset in RA and Dec from all
VLA observations of GRB\,980703 (small cross).  The larger
cross is the offset measurement from \citep{bkd01}.  The solid circle
designates the  projected maximum source size
from the radio observations, and the dashed circle is the optical size
from \citep{hfh+01}.  Clearly the formation of massive stars is
concentrated in the central region of the host, and the small offset
of the burst from the host center indicates that GRB\,980703 occurred
in the region of maximum star formation \cite{bkf01}.  This points to
a link between GRBs and massive stars.}
\label{fig:offset}
\end{figure}

\begin{figure}
\resizebox{0.48\textwidth}{!}{\includegraphics{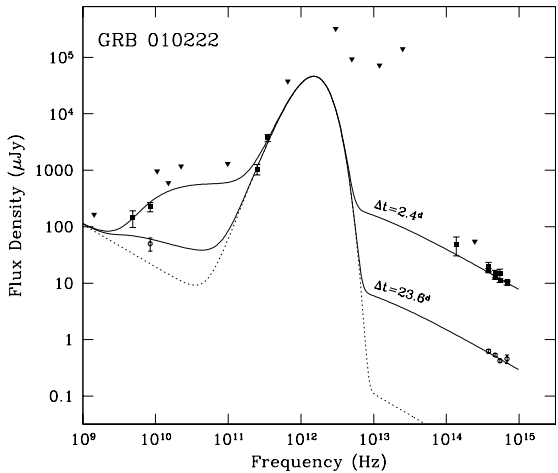}}
\caption{Spectral energy distribution of the host+afterglow emission
from GRB\,010222.  The dashed line is the contribution of the host,
and the solid line is the fading afterglow.  The sub-mm flux densities
at 250 and 350 GHz point to SFR$\sim 750$ M$_\odot$/yr, and a ULIRG
host galaxy.  From \cite{fbm+01}}
\label{fig:010222sp}
\end{figure}

\end{section}

\begin{section}{Future Prospects}

Future radio/sub-mm/FIR studies of GRB hosts will be augmented by X-ray
observations in order to assess the importance of obscured AGN in
these host galaxies.  There is some indication from studies of local
ULIRGs \cite{sm96} that high SFR is usually accompanied by some AGN
activity.  Thus, the advent of observatories such as the EVLA,
SKA, SIRTF, and ALMA, in addition to XMM and Chandra will
greatly increase our ability to study the properties of these hosts
with greater sensitivity and resolution.  For example, with a factor
ten increase in resolution and a factor five increase in sensitivity
over the current VLA, we will be able to probe scales of approximately
5 mas with the EVLA; for a galaxy at $z\sim 1$ this translates to a
physical scale of 150 pc.  In addition, EVLA will detect galaxies with
a total SFR as low as 50 M$_\odot$/yr at $z\sim 1$.  

The potential of a GRB-selected galaxy sample is immense and unique.
The dust-penetrating power of GRBs and their broad-band afterglow
emission, offer a number of unique diagnostics: the obscured star
formation fraction, the ISM within the disk, the local environment of
the burst, and global and line-of-sight extinction, to name a few.  In 
addition, GRBs allow us to select a wide range of galaxies independent 
of their emission propertied in any wavelength regime, and in addition 
they supply redshift information for these galaxies; the lack of
accurate redshifts is one of the main problems of sub-mm studies of
high-redshift galaxies. 

It appears, therefore, that the numerous and detailed optical studies
of GRB hosts are only the tip of the iceberg in our understanding of
galaxies at high redshifts.  

\end{section}


\begin{thebibliography}{10}

\bibitem{mfd+96}
{Madau}, P., et al. 1996, MNRAS, 283, 1388.

\bibitem{rauch98}
{Rauch}, M. 1998, ARAA, 36, 267.

\bibitem{sw00}
{Storrie-Lombardi}, L.~J. \& {Wolfe}, A.~M. 2000, ApJ, 543, 552.

\bibitem{as00}
{Adelberger}, K.~L. \& {Steidel}, C.~C. 2000, ApJ, 544, 218.

\bibitem{rkf+98}
Ramaprakash, A.~N., et al. 1998, Nature, 393, 43.

\bibitem{bcr00}
{Barger}, A.~J., {Cowie}, L.~L., \& {Richards}, E.~A. 2000, AJ, 119, 2092.

\bibitem{prb+00}
{Peacock}, J.~A., et al. 2000, MNRAS, 318, 535.

\bibitem{bkd01}
Bloom, J.~S., Kulkarni, S.~R., \& Djorgovski, S.~G. 2001, submitted to
AJ; astro-ph/0010176.

\bibitem{lr00}
{Lamb}, D.~Q. \& {Reichart}, D.~E. 2000, ApJ, 536, 1.

\bibitem{afv+00}
{Amati}, L., et al. 2000, Science, 290, 953.

\bibitem{pgg+00}
{Piro}, L., et al. 2000, Science, 290, 955.

\bibitem{bkf01}
Berger, E., Kulkarni, S. R., \& Frail, D. A. 2001, ApJ, 560, 652.

\bibitem{fbm+01}
Frail, D. A., et al. 2001, Accepted to ApJ; astro-ph/0108436.

\bibitem{rei01}
Reichart, D.~E. 2001,  Submitted to ApJL; astro-ph/0107546.

\bibitem{c92}
Condon, J.~J. 1992, ARAA, 30, 575.

\bibitem{cy99}
{Carilli}, C.~L. \& {Yun}, M.~S. 1999, ApJ, 513, L13.

\bibitem{cy00}
{Carilli}, C.~L. \& {Yun}, M.~S. 2000, ApJ, 530, 618.

\bibitem{dce00}
Dunne, L., Clements, D.~L., \& Eales, S.~A. 2000, MNRAS, 319, 813.

\bibitem{sm96}
Sanders, D.~B. \& Mirabel, I.~F. 1996, ARAA, 34, 749+.

\bibitem{hpw+00}
Haarsma, D.~B., et al.  2000, ApJ, 544, 641.

\bibitem{dkb+98}
Djorgovski, S.~G., et al.  1998, ApJ, 508, L17.

\bibitem{rfk+99}
Richards, E.~A., et al.  1999, ApJ, 526, L73.

\bibitem{r00}
Richards, E.~A. 2000, PASP, 112, 1001.

\bibitem[Rola, Terlevich and Terlevich(1997)]{rtt97}
Rola, C.~S., Terlevich, E., \& Terlevich, R.~J. 1997, MNRAS, 289, 419.

\bibitem{k92}
Kennicutt, R.~C. 1992, ApJ, 388, 310.

\bibitem[Holland et al.(2001)]{hfh+01}
Holland, S., et al.  2001, submitted to A\&A. astro-ph/0103058.


\end{thebibliography}
\end{document}